# Analysis of Trajectory Similarity and Configuration Similarity in On-the-Fly Surface-Hopping Simulation on Multi-Channel Nonadiabatic Photoisomerization Dynamics

Xusong Li,[1,2,3] Deping Hu,[1,3] Yu Xie,[1] and Zhenggang Lan[1,2,a)]

[1]CAS Key Laboratory of Biobased Materials, Qingdao Institute of Bioenergy and Bioprocess Technology, Chinese Academy of Sciences, Qingdao 266101, China

[2]Sino-Danish Center for Education and Research/Sino-Danish College, University of Chinese Academy of Sciences, Beijing 100049, China

[3]University of Chinese Academy of Sciences, Beijing 100049, China

**ABSTRACT**

We propose an "automatic" approach to analyze the results of the on-the-fly trajectory surface hopping simulation on the multi-channel nonadiabatic photoisomerization dynamics by considering the trajectory similarity and the configuration similarity. We choose a representative system phytochromobilin chromophore model (PΦB) to illustrate the analysis protocol. After a large number of trajectories are obtained, it is possible to define the similarity of different trajectories by the Fréchet distance and to employ the trajectory clustering analysis to divide all trajectories into several clusters. Each cluster in principle represents a photoinduced isomerization reaction channel. This idea provides an effective approach to understand the branching ratio of the multi-channel photoisomerization dynamics. For each cluster the dimensionality reduction is employed to understand the configuration similarity in the trajectory propagation, which provides the understanding of the major geometry evolution features in each reaction channel. The results show that this analysis protocol not only assigns all trajectories into different photoisomerization reaction channels, but also extracts the major molecular motion without the requirement of the pre-known knowledge of the active photoisomerization site. As a side product of this analysis tool, we can also easily to find the so-called "typical" or "representative" trajectory for each reaction channel.

## 1. INTRODUCTION

Photoinduced isomerization reactions via the double-bond twisting motions on molecular excited states widely exist in photochemistry.[1-4] For instance the photoisomerization processes of the chromophores in photoreceptor proteins are the primary steps in the solar-to-mechanical energy conversions, which trigger important photoinduced biological functions.[1,2,4-6] The photoisomerization mechanism received considerable research interests in last decades.[1-5,7,8] Among these studies, theoretical calculations clarified that nonadiabatic dynamics at conical intersections are essential for photoisomerization processes.[1,2] The simulation of nonadiabatic dynamics needs to take the coupled electron-nucleus motion into account, in which Born-Oppenheimer approximation breaks down.[9,10] Although many theoretical approaches were proposed to solve nonadiabatic dynamics,[2,3,9-41] trajectory surface hopping (TSH) approaches become popular due to their simplicity and easy implementation.[32,34,40-54] With the development of computational facilities, the on-the-fly TSH dynamics provides us a reasonable way to simulate the nonadiabatic dynamics of polyatomic molecules by inclusion of all degrees of freedom.[7,34,41-47,51,52,55-68] Nowadays the combination of the on-the-fly dynamics and TSH (or other theoretical approaches) becomes a promising tool to understand the photoisomerization mechanism at the atomic level.[1-3,7,8,41-43,45-47,51,60,63,65,69-76]

The on-the-fly TSH dynamics often requires the computation of a large number of trajectories. The statistical analysis over all trajectories gives various dynamical features, for instance, the excited-state population decay, the structure evolution and the geometrical features at PES crossings. In the typical analysis of the TSH results, the active reaction coordinates are normally identified by the eye view of many trajectories and the results are discussed by the explanation of a few "representative" trajectories.[42,43] This approach also largely relies on the preliminary understanding of the nonadiabatic dynamics, such as the reaction pathways and the relevant conical intersections. This "eye-view" analysis routine becomes not an easy task, when the system size becomes large, the complicated molecular motions are involved, many trajectories are concerned, or the pre-known knowledge on the reaction channels is missing. Thus, the novel analysis tool should be developed to examine the TSH simulation results, particularly because more and more studies take the on-the-fly TSH calculations to treat different nonadiabatic dynamics of various complicated systems.[41] As a typical example, the analysis of the TSH simulation on the photoisomerization dynamics is not trivial, because the twisting motions may happen at different twisting sites, the major motion may involve the strong couples between different nuclear degrees of freedom, and several reaction channels may result in different photoproducts.

Unsupervised machine learning algorithms, particularly dimensionality reduction approaches, such as principle component analysis (PCA),[77-79] multidimensional scaling (MDS),[80,81] isometric feature mapping (ISOMAP),[82,83] diffusion map,[84,85] autoencoder[86] etc., were employed to examine the main feature of the geometrical evolution in the ground-state molecular dynamics simulation.[87-94] In recent years some groups tried to use such tools in the analysis of nonadiabatic dynamics,[95-100] which tried to automatically extract the main geometrical feature of the trajectory evolution. The underlining idea is as following. A single geometry in a trajectory is represented by a point in a high-dimensional coordinate space. After the collection of a large number of geometries generated by the trajectory propagation in the nonadiabatic dynamics, these unsupervised ML approaches construct a mapping from the high dimensional space to a low dimensional space, which tries to conserve the pattern feature of data point distribution. The active motion responsible for the nonadiabatic dynamics was then examined in the low-dimensional space. These efforts help us to understand the geometric evolution in the nonadiabatic dynamics. However, the application of these approaches in real analysis tasks may not be fully straightforward. For instance, such idea may not work properly in the multi-channel situations, because different reactive coordinates may be responsible for different channels. Most importantly, the analysis in the configuration space does not directly take an important dynamic feature namely "time evolution" into account. Instead the time feature is indirectly included afterwards, through monitoring the movement of the data set in the low-dimensional space constructed by the dimensionality reduction.

In this paper, we propose an improved "automatic" approach to analyze the on-the-fly TSH results by re-considering the concept of "trajectory evolution with time being". Instead of only performing the dimensionality reduction in coordinate space, we also examine the trajectory evolution in the so-called "trajectory space", in which we measure the "distance" or "dissimilarity" between different trajectories. The estimation of the trajectory similarity is widely employed in various scientific fields.[101-109] In the current work the so-called Fréchet distance[109-112] was taken to evaluate the "dissimilarity" between two trajectories. After the construction of the pair-wise dissimilarity matrix for all trajectories, the clustering method is employed to assign the trajectories into different groups. In this trajectory clustering analysis, each group in principle should represent a reaction channel. The reactive coordinate responsible for each channel is furtherly identified by the dimensionality reduction approaches in the coordinate space, suggested by our previous work.[95] In one word, this analysis way considers first the trajectory similarity and second the configuration similarity, which makes the analysis procedure more transparent and automatically. This provides us a powerful tool to analyze the nonadiabatic dynamics with many reactive channels.

As the first attempt, we wish to know whether the above idea can clearly identify distinguishing channels and clarify their active motion in the photoisomerization dynamics. The reason is the photoisomerization serves a kind of prototype reactions, in which the twisting motions at different sites give rather different reaction channels and several distinguishing photoproducts are formed as the result.[1-4] Thus, in principle, this type of the nonadiabatic dynamics provided us a very good model to examine our idea on the estimation of the trajectory similarity and the geometrical similarity. In this work, we take the photoisomerization of the phytochromobilin (PΦB in Fig 1) model as an example to check the performance of the above proposed analysis method. As widely-existing plant's photoreceptors, the PΦB and other phytochromes were studied extensively.[1,6,70,113-122] The PΦB system decays to the ground state via different conical intersections and finally several photo-products are formed.[1,70,122] Thus, the PΦB model is an ideal system to test our new approach. The results show that the analysis approach with the combination of the trajectory similarity and the configuration similarity is a very powerful protocol that can perform the automatic and efficient analysis of nonadiabatic photoisomerization dynamics with several channels and different products. Although the current work is based on the TSH calculations of photoisomerization, it is also possible to use the similar idea to understand other types of trajectory-based nonadiabatic dynamics simulation.

This work is organized as follows. Section 2 outlines theoretical methods, implementation and computational details. Section 3 shows results and Section 4 performs discussions. Section 5 gives the conclusion of the current work.

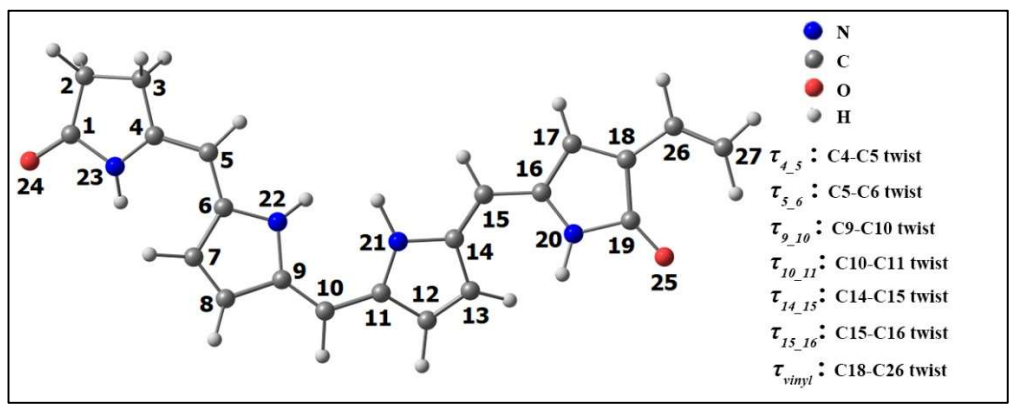

**Fig 1**. The model of the ZaZsZa isomer of PΦB and some key coordinates.

## 2. THEORETICAL METHODS AND COMPUTATIONAL DETAILS

### 2.1 Theoretical Methods

#### 2.1.1 Trajectory Surface Hopping Dynamics

Many previous works have provided the detailed discussion on Tully's TSH approaches,[32,42,43,46,47,52] so we outline the main concept here. In the Tully's TSH framework, the nuclear motion was treated by the classical Newtonian mechanics, while the electronic motion was described by the quantum evolution. The nonadiabatic transitions were described by the trajectory hops between different electronic states and the hopping probability was determined by Tully's fewest switches algorithm.[32] The initial conditions (such as geometries and velocities) were sampled by the Wigner distribution of the ground vibrational level of the normal modes on the electronic ground state.

#### 2.1.2 Configuration Similarity Definition

Mathematically, a single geometry is represented by a point in a high-dimensional coordinate space, which is characterized by a high-dimensional vector. Thus, the similarity/dissimilarity between two geometry snapshots is measured by the distance between two corresponding points in the metric view. This provides us the basic idea on the definition of the dissimilarity matrix **D** over all snapshots. Following previous work,[95-97] the elements $d_{ij}$ in the **D** matrix was defined by the root mean square distance (RMSD) of two configurations. In the RMSD calculations, it is necessary to

remove the contribution of translational and rotational motions.[95,123,124]

### 2.1.3 Trajectories Similarity Definition

One of central ideas in this work involves the definition of the similarity between two trajectories. Different numerical approaches were proposed to compute the trajectory similarity, such as Hausdorff distance, Fréchet distance and so on.[102,111] Among them, the Fréchet distance is a good candidate to conduct the analysis of trajectory evolution, because the chronological order is taken in to account explicitly and this analysis approach also does not require the same propagation duration for all trajectories. [102,103,111]

Roughly speaking, it is possible to understand the Fréchet distance in an intuitive way. Let us assume that a man is walking along a path **P** and his dog is running along another path **Q**. They are connected by a leash in the whole walking procedure. Both starting and ending points are known for path **P** and path **Q**. The man and his dog move along their own pathways independently under constrain that their motion must follow the monotonic chronological way from the starting point to the ending point, and no backwards movement is allowed. When the dog changes speed to make the leash as slack as possible, the length of the shortest leash sufficient for both man and his dog moving along their own paths defines the Fréchet distance between two curves **P** and **Q**.

Next, we discussed the formal mathematical view of the Fréchet distance.[109-112] Suppose that **P** and **Q** are two given curves in the metric space $V_s$, which are represented by the continuous mappings as below

$$\begin{aligned}\mathbf{P}: & [p_0, p_1] \to V_s \quad [p_0, p_1 \in \mathbf{R}_s, \quad p_0 \leq p_1]\\ \mathbf{Q}: & [q_0, q_1] \to V_s \quad [q_0, q_1 \in \mathbf{R}_s, \quad q_0 \leq q_1]\end{aligned} \quad (1)$$

where $p_0, p_1$ (or $q_0, q_1$) are starting and ending points of the curve **P** (or **Q**) in the space of $\mathbf{R}_s$. The Fréchet distance between **P** and **Q** is defined as

$$\delta_F(\mathbf{P},\mathbf{Q}) = \inf_{\alpha,\beta} \max_{t\in[0,1]} \{dist[\mathbf{P}(\alpha(t)),\mathbf{Q}(\beta(t))]\} \quad (2)$$

Where $\alpha(t)$ (or $\beta(t)$) is an arbitrary continuous non-decreasing function that maps the unit interval [0,1] onto [$p_0, p_1$] (or [$q_0, q_1$]), namely $\alpha(0) = p_0$ and $\alpha(1) = p_1$ (or $\beta(0) =$

$q_0$, $\beta(1) = q_1$).

For computational practices, an arbitrary continuous curve is typically approximated by a polygonal curve, and thus the discrete Fréchet distance,[109-112] instead of its continuous counterpart, is often used to examine the dissimilarity of two polygonal curves.

Two trajectories **P** and **Q** are approximated by polygonal curves represented by two sequences **S(P)** ($p_1$, …$p_i$…, $p_n$) and **S(Q)** ($q_1$, …$q_j$…, $q_m$), where $p_i$ is the i[th] snapshot of the trajectory **P** and $q_j$ is the j[th] snapshot in the trajectory **Q**. The coupling **C** between **P** and **Q** in the production space **S(P)** × **S(Q)** is given by a sequence

$$\mathbf{C(P,Q)} \equiv (p_{a_1}, q_{b_1}), (p_{a_2}, q_{b_2}), \ldots, (p_{a_k}, q_{b_k}), \ldots, (p_{a_T}, q_{b_T}) \quad (3)$$

with correct starting and ending conditions $a_1 = b_1 = 1$, $a_T = n$, $b_T = m$.

Notice that here the lengths of **S(P)** and **S(Q)**, namely n and m, may not be the same. However, it is always possible to construct the **C(P, Q)** because two successive elements, for instance $p_{a_k}$ and $p_{a_{k+1}}$, may be the same. More precisely, starting from a point pair ($p_{a_k}$, $q_{b_k}$), one point (or both points) should move to its next position (or their next positions) at each step. This means that one of the below three conditions should be satisfied:

$$\begin{aligned}
(I) \quad & a_{k+1} = a_k + 1 \quad & b_{k+1} = b_k \\
(II) \quad & a_{k+1} = a_k \quad & b_{k+1} = b_k + 1 \\
(III) \quad & a_{k+1} = a_k + 1 \quad & b_{k+1} = b_k + 1
\end{aligned} \quad (4)$$

When the coupling **C** is calculated, the corresponding coupling distance is defined as the largest distance between $p_{a_k}$ and $q_{b_k}$.

$$\|\mathbf{D_C}\| \equiv \max_{k=1,2,\ldots,T} dist(p_{a_k}, q_{b_k}) \quad (5)$$

Because the coupling between two given trajectories **P** and **Q** is not uniquely defined, all possible couplings **C** form a space $R_s(\mathbf{C})$. The discrete Fréchet distance between **P** and **Q** is defined as the minimum coupling distance over all possible couplings in the space $R_s(\mathbf{C})$, namely

$$\delta_{dF} \equiv \min\{ \|\mathbf{D_c}\| \mid \mathbf{C} \in R_s(\mathbf{C}) \} \tag{6}$$

According to this idea, it is possible to compute the discrete Fréchet distance by the dynamical programming algorithm.[109,110,112,125] This allows us to compute the pair-wise dissimilarity matrix over all trajectories, giving the possibility to employ various machine learning algorithms in the further analysis.

**2.1.4 Multi-Dimensional Scaling**

As a widely-used dimensionality reduction method, the classical MDS constructs the low-dimension space, in which the pair-wise dissimilarities between all data points under study are preserved.[80] The MDS algorithm starts from the construction of the pair-wise dissimilarity matrix **D** with the dimension n × n, n is the number of objects. $d_{ij}$ resents the "distance" between two data points, and then it is possible to define the scalar product matrix **B** as

$$\mathbf{B} = -\frac{1}{2}\mathbf{J}\mathbf{D}^{(2)}\mathbf{J} \tag{7}$$

Where $\mathbf{D}^{(2)}$ is the squared proximity matrix with elements $d_{ij}^2$, namely

$$\mathbf{D}^{(2)} = \left[ d_{ij}^{2} \right] \tag{8}$$

and **J** is the center matrix define as

$$\mathbf{J} = \mathbf{I} - n^{-1}\mathbf{1}\mathbf{1}^{\mathrm{T}} \tag{9}$$

where **I** is an unit matrix. **1** is a column vector with all elements equal to 1 and $\mathbf{1}^{\mathrm{T}}$ is the corresponding row vector. Thus, the product of these two matrices $\mathbf{1}\mathbf{1}^{\mathrm{T}}$ gives a matrix with all elements equal to 1.

Next, we diagonalize the **B** matrix and reorder all eigenvalues from largest to smallest. The larger eigenvalue corresponds to more important dimension. For instance, if a reduced space with *m*-dimension is considered, we need to take the *m* largest positive eigenvalues $\lambda_1...\lambda_m$ and their corresponding eigenvectors $e_1...e_m$. The coordinates of

all data points in the low-dimensional space are computed by

$$\mathbf{L} = (e_1 \cdots e_m) \begin{pmatrix} \sqrt{\lambda_1} & \cdots & 0 \\ \vdots & \ddots & \vdots \\ 0 & \cdots & \sqrt{\lambda_m} \end{pmatrix} \quad (10)$$

The relative embedding error is computed by the stress unction, see the MDS textbook.[80]

**2.1.5 DBSCAN Clustering Algorithm**

Here we selected the DBSCAN (density-based spatial clustering of applications with noise)[126,127] algorithm to perform the trajectory clustering analysis after the construction of the pair-wise dissimilarity matrix of all trajectories. The DBSCAN is a density-based clustering algorithm. The basic assumption of the DBSCAN method is that all data points form the high-density and low-density areas. Then it is possible to put all points belonging to the same high-density area together to define a cluster, while different clusters are separated by the low-density areas. The data points located in the low-density areas are labelled as outliers. Because a large number of trajectories are computed in the TSH calculation, it is possible to get a few abnormal trajectories, which give a few so-called outer data points that do not belong to any cluster in the trajectory clustering analysis. As a density-based method, the DBSCAN cluster algorithm is robust to outlier points relevant to these abnormal trajectories.

**2.2 Implementation Details**

Next, we discuss all implementation details by taking the nonadiabatic dynamics of the PΦB model photoisomerization as an example.

**2.2.1 On-the-Fly TSH Dynamics Simulation**

The analysis protocol certainly starts from the on-the-fly TSH calculations. The nonadiabatic photoisomerization dynamics of the PΦB model is investigated by the TSH method at the semiempirical OM2/MRCI level.[128-130] All trajectories start from the $S_1$ state and the propagation lasts up to 1 ps. We use the same computational setups, such as the active space, discussed in previous works[70,95,96]. To analyze the simulation data easily, the TSH calculations are performed by the JADE code[46] by calling the OM2/MRCI [128-130] calculations with MNDO code [131]. In the current work, the trajectory clustering analysis requires a large number of trajectories (see the next section of results).

**2.2.2 Analysis of Excited-State Photoisomerization Dynamics before $S_1$-$S_0$ Hops**

Normally the analysis of the multi-channel nonadiabatic dynamics should identify which conical intersection is responsible for the internal conversion and which molecular motion is relevant to the excited-state dynamics. For the current PΦB model, this task becomes the identification of the different isomerization channels via different conical intersections. To address these key questions, the following protocols are employed.

(a) We selected the geometries at every 10 fs for each trajectory before the $S_1$-$S_0$ hops. In this sense, the excited-state dynamics before the $S_1$ decay is fully characterized by these trajectories containing a large number of snapshots.

(b) For two trajectories **P** and **Q**, we computed the distances between any two geometries $p_i$ ($p_i \in$ **P**) and $q_j$ ($q_j \in$ **Q**). In this step, the distance between two geometries are defined by their RMSD with neglecting hydrogen atoms. We performed the alignment of each snapshot with respect to the reference geometry (ground-state minimum) to remove the contribution of translational and rotational motion. This alignment approach, instead of the pair-wise alignments for all snapshots, confirms that a correct metric space is formed in the Fréchet distance calculations.[109,125,132]

(c) The dissimilarity of each pair of trajectories is defined by the discrete Fréchet distance. Finally, we got the pairwise distance matrix $\mathbf{D}_{traj}$ of all trajectories, whose dimension is $N_{traj} \times N_{traj}$.

(d) The MDS analysis was performed in the basis of the pairwise distance matrix $\mathbf{D}_{traj}$ of all trajectories. Then in the two-dimensional space each trajectory is represented by a point and the basic feature of the data distribution is easily examined. When two data points are closer, two corresponding trajectories are more "similar".

(e) The trajectory clustering analysis was performed with the DBSCAN clustering

algorithm, which divide all data points to different groups in the two-dimensional space. In the trajectory clustering, the trajectories with high similarity in principle should be assigned into the same group.

(f) In the ideal case each cluster corresponds to a decay channel in the nonadiabatic photoisomerization dynamics after the trajectory clustering analysis. For this purpose, we performed the additional check. The clustering analysis divided all trajectories into different groups. Next based on the trajectories belonging to the same cluster (for instance Cluster **A**), we took their Fréchet distances to construct the pair-wise distance matrix (labelled as $\mathbf{D}_{traj\_A}$), which is the submatrix of the full pair-wise dissimilarity matrix $\mathbf{D}_{traj}$ of all trajectories. Based on $\mathbf{D}_{traj\_A}$, the MDS dimensionality reduction and the DBSCAN clustering algorithm were performed again, to see whether it is possible to divide Cluster **A** into several smaller sub-clusters. Notice that the different reduced spaces were formed at two successive runs, because the different distance matrices were employed in the dimensionality reduction before the clustering analysis. This procedure should be repeated until each generated small cluster only gives a single dense data set. Until now, we wish that each cluster in principle corresponds to a single nonadiabatic decay channel.

(g) The next task is to identify which reactive coordinates are responsible for a single nonadiabatic decay channel. In this step, we simply took the dimensionality reduction analysis discussed in our previous work.[95] The trajectories belonging to the same cluster were collected. All snapshots belonging to the selected trajectories were used to calculate the pair-wise dissimilarity matrix $\mathbf{D}_{geom}$. Then the MDS analysis based on $\mathbf{D}_{geom}$ was performed to construct the low dimensional space, and each point now refers to a configuration. For the data points located in the same grid area, we overlapped their configurations together and examine the characteristic geometric feature. In this way, it is possible to identify the major reactive coordinate responsible for a particular channel in the nonadiabatic decay dynamics.

**2.2.3 Analysis of Full Nonadiabatic Photoisomerization Dynamics towards Different Photoproducts**

We collected trajectories belonging to the same cluster (for instance Cluster **A**)

generated from the analysis of excited-state dynamics before $S_1$-$S_0$ hops, after making sure that each cluster should not be divided again. These trajectories in principle pass the same conical intersections, while different products may be formed after internal conversion. Next, we wish to understand their full nonadiabatic dynamics towards photoproducts. We expect that Cluster **A** can be divided into several smaller clusters again after the ground-state dynamics is considered.

The geometry re-sampling for these trajectories is performed with a larger time step (40 fs) and a longer time duration (1 ps). The employment of the longer time duration confirms that photoproducts are formed by the successive ground-state dynamics after the internal conversion. The use of the larger time step is mainly for reducing computational cost. We performed the trajectory clustering analysis again by taking the ground-state dynamics into account for the trajectories passing the same conical intersection. Several clusters were formed, and we hope that each cluster includes trajectories with high similarity, namely passing the same conical intersection and forming the same photoproduct. The analysis of the geometry similarity with the dimensionality reduction approach is again employed for each group of trajectories, to clarify the major molecular motions in a channel.

### 2.2.4 The Definition of the "Typical Trajectory"

As a side product of trajectories clustering process, it is easy to find the "typical trajectory" for each reaction channel. As discussed in the previous section, all trajectories belonging to the same non-dividable cluster should be "similar" in trajectory clustering analysis. Thus, if one trajectory shows the highest similarity with all rest trajectories within a cluster, this one can be assigned as the "typical" trajectory.

Starting from all trajectories belonging to the same cluster, we estimated their similarity via the pair-wise Fréchet distance matrix. Among all trajectory, it is always possible to find a trajectory, which gives the minimum value of the sum of the Fréchet distances between this selected trajectory and all other trajectories. In this situation, we can assign this trajectory as the "typical" or "representative" one that characterizes the important geometrical evolution of this group of trajectories.

**2.3 Coding Issues**

In this work the dynamics simulation was done within the developing version of the JADE package,[46,47,133] which contains a module to interface with several quantum chemistry packages (including the interface with the MNDO package[131]). A simple homemade FORTRAN code was developed to calculate the RMSD between two geometries.[95] Most analysis scripts were written with Python language and the Scikit-learn Python toolkit[134,135] was used for the data analysis, such as the DBSCAN clustering.

## 3. RESULTS

**3.1 Clustering Analysis of Trajectory Similarity before $S_1$-$S_0$ Hops**

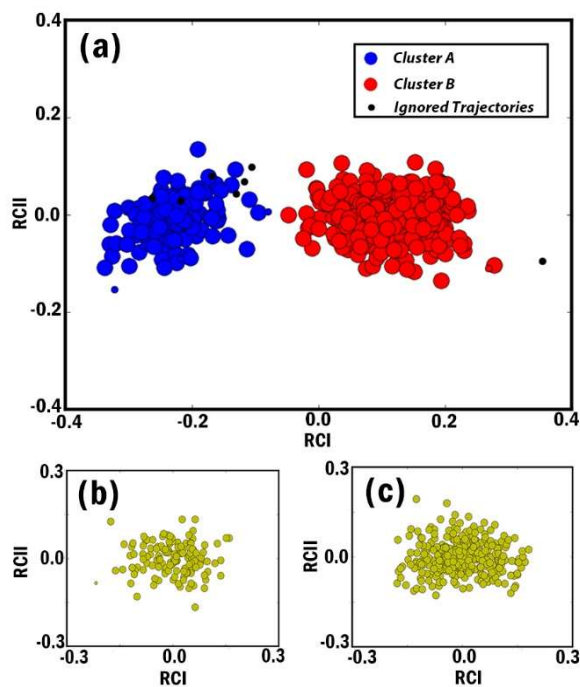

**Fig 2.** The clustering analysis of trajectory similarity before $S_1$-$S_0$ hops. (a) In the first run, we simply collected all trajectories, defined the pair-wise distance matrix for all trajectories, employed the dimensionality reduction approach and performed the clustering analysis. Two clusters appear, labelled as Cluster **A** and Cluster **B**. (b) In the second run, we took all trajectories belonging to Cluster **A** and repeated the above analysis as the first step. (c) The similarity analysis was also performed for trajectories belonging to Cluster **B**.

In the analysis of nonadiabatic dynamics of photoisomerization, an important task is to understand the excited-state dynamics before the $S_1$-$S_0$ decay. Thus, we cut the trajectories until their hops, defined the pair-wise distance matrix among all trajectories by invoking the Fréchet distance calculations, used the dimensionality reduction approach by the MDS and performed the trajectory clustering analysis with DBSCAN methods. Two clusters appear clearly (Figure 2 (a)), which are labelled as Cluster **A** and Cluster **B**.

Cluster **A** contains 142 trajectories and Cluster **B** contains 303 trajectories, while a few trajectories (~ 2.8%) were ignored according to the noise reduction principle of the DBSCAN algorithm. Although only two clusters exist, it is necessary to check whether each cluster can be divided again. For this purpose, at the second step we took all trajectories belonging to Cluster **A**, defined the pair-wise Fréchet distance matrix, performed the dimensionality reduction approach and the trajectory clustering analysis again. Fig 2 (b) shows that it is not possible to divide Cluster **A** into small groups. The same operation on Cluster **B** was performed and the results are given in Fig 2 (c). We wish to point out that at each step different reduced spaces are formed, because different distance matrixes were employed in the MDS analysis. When each cluster couldn't be divided anymore after several iterative steps of trajectory clustering analysis, two clusters are finally obtained. In principle, this indicates that two nonadiabatic decay channels may be involved. The next task is to check the trajectory feature of each group and to understand the dynamical evolution in each channel.

### 3.1.1 Geometrical Evolution of Trajectories belonging to Cluster A

To get the geometrical features of trajectory belonging to Cluster **A**, we first collected all trajectories belonging to such cluster. The snapshot was taken before the hopping events and totally 2827 geometries were collected to form a data set.

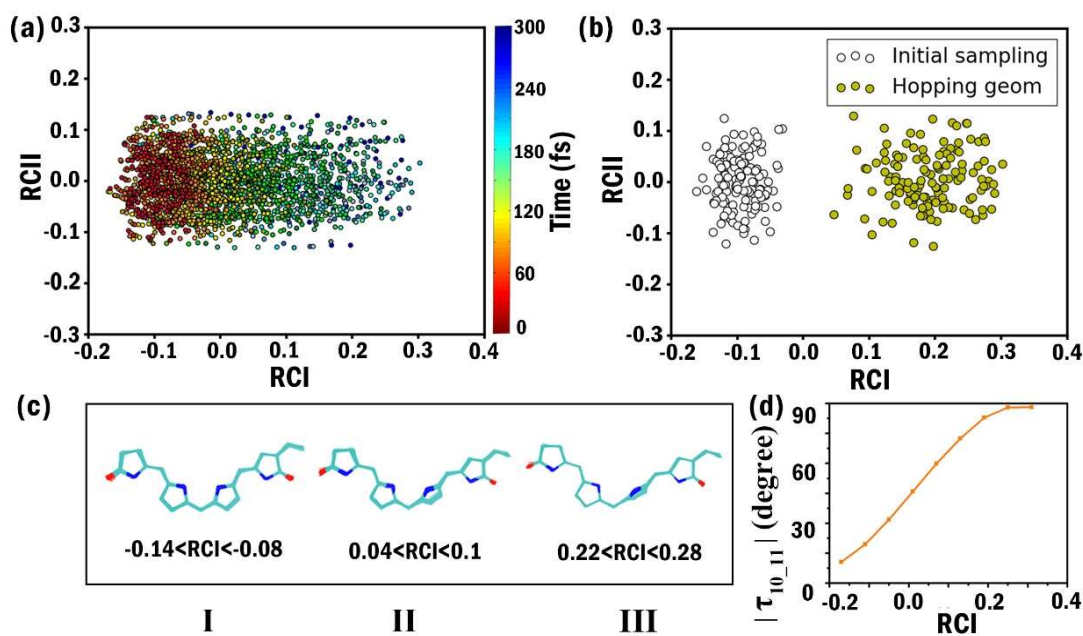

**Fig 3.** Classical MDS analysis of the geometrical evolution for the trajectories belonging to Cluster **A**. (a) Locations of sampled geometries in the low-dimensional space spanned by two leading reduced coordinates RCI and RCII. Colour codes indicate the time evolution. (b) Locations of the initial geometries and the hopping geometries in the two-dimensional reduced space. (c) Geometrical aggregations in three representative local domains. (d) The values of $\tau_{10\_11}$ vs RCI.

This classical MDS analysis of the pair-wise distance matrix among all collected geometries gives a clear distribution pattern in the low-dimensional space spanned by two reduced coordinates as shown in Fig 3 (a), in which each point represents a geometry snapshot. It's obvious that the snapshots evolve from the Franck-Condon (FC) region to the $S_1$-$S_0$ conical intersection region, corresponding to the changing of RCI values from ~ -0.1 to ~ 0.2 as shown in Fig 3 (b). We selected three representative local domains along the RCI axis and stacked all snapshots in each selected domain. It turns out that the RCI was governed by the torsional angle at the $C_{10}$-$C_{11}$ bond, namely $\tau_{10\_11}$, as shown in Fig 3 (c) and (d). Overall, the torsional motion of $\tau_{10\_11}$ is observed and the hops take place near the $S_0$-$S_1$ conical intersection region with $\tau_{10\_11}$ ~ 70°-90°, as shown in Fig 3 (b) and (d).

### 3.1.2 Geometrical Evolution of Trajectories belonging to Cluster B

Similar analysis was also performed for Cluster **B**. We totally collected 6379 geometries for MDS analysis. As shown in Fig 4 (a), the dominant reaction coordinates of the trajectories in Cluster **B** can also be represented by one-dimensional reduced coordinate (RCI), see Fig 4 (b). The torsional motion at the $C_9$-$C_{10}$ bond ($\tau_{9\_10}$) is responsible for RCI as shown in Fig 4 (c) and (d). This indicates that the strong torsional motion $\tau_{9\_10}$ take place in the excited-state decay pathway towards to conical intersections.

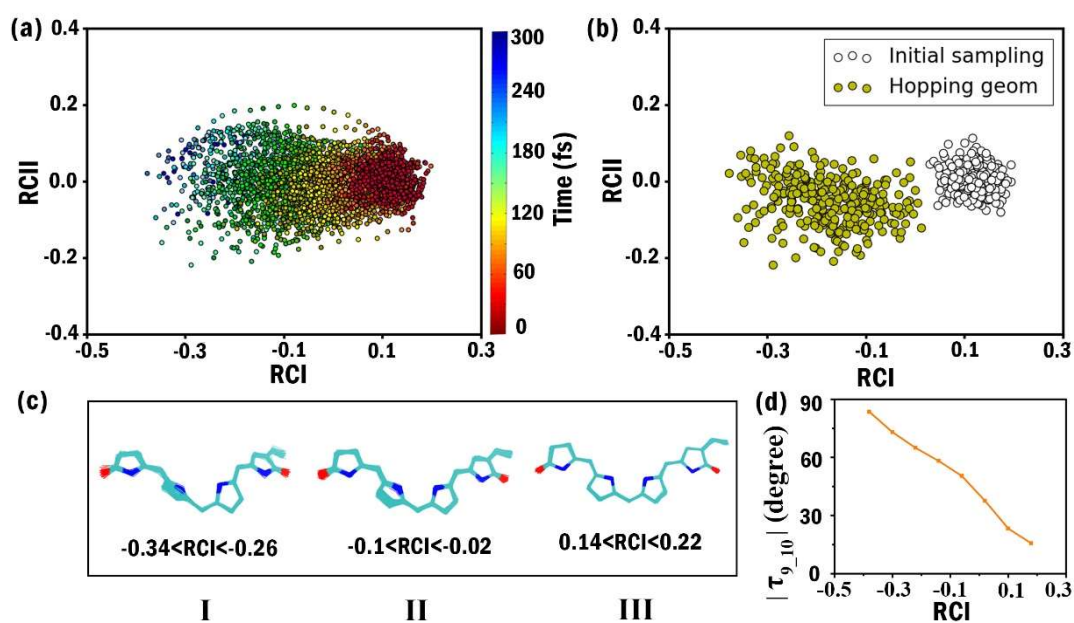

**Fig 4.** Classical MDS analysis of the geometrical evolution for the trajectories belonging to Cluster **B**. (a) Locations of sampled geometries in the low-dimensional space spanned by two leading reduced coordinates RCI and RCII. Colour codes indicate the time evolution. (b) Locations of the initial geometries and the hopping geometries in the two-dimensional reduced space. (c) Geometrical aggregations in three representative local domains. (d) The values of $\tau_{9\_10}$ vs RCI.

In the above protocol, two clusters are formed in the clustering analysis of the trajectory similarity, the further MDS analysis of the geometry similarity shows that each cluster corresponds to a single reaction channel. It's obvious that Cluster **A** is relevant to the torsional motion along $\tau_{10\_11}$ while Cluster **B** is governed by the torsional motion of $\tau_{9\_10}$. The above detailed analysis gives a clear description of the dynamics process from the initial sampling to two $S_0$-$S_1$ intersection regions. These observations are consistent with our previous studies.[70]

If we wish to get a full dynamical picture of photoinduced reactions, it is also important to know photoproducts. We repeated the above analyses of trajectory similarity and geometry similarity again, while this time all trajectories stop at 1 ps. For all trajectories belonging to either Cluster **A** or **B**, we computed the trajectory similarity and perform the clustering analysis again, while the ground-state dynamics after the internal conversion is also included. This way clearly shows that the trajectories of Cluster **A** (or **B**) can be distinguished by their different photoproducts. Because many analysis procedures are very similar, we tried to mainly focus on the analysis of Cluster B containing more trajectories in the below illustration. To avoid redundancy, we gave the main results relevant to Cluster **A** in Supporting Information.

**3.2 Clustering Analysis of Trajectory Similarity with the Inclusion of Photoproducts**

After the inclusion of the ground state dynamics, Cluster **B** is divided into two sub-clusters as shown in Fig 5 (a), which are labelled as Cluster **B1** and **B2**. Cluster **B1** contains 191 trajectories, while Cluster **B2** contains 106 trajectories, while a few trajectories are treated as noise in the DBSCAN algorithm. The second round of trajectories clustering results as shown in Fig 5 (b) and (c) prove that these two sub-clusters can't be divided anymore. The appearance of two clusters, **B1** and **B2**, indicates that two different products are formed for the trajectories passing the same conical intersection.

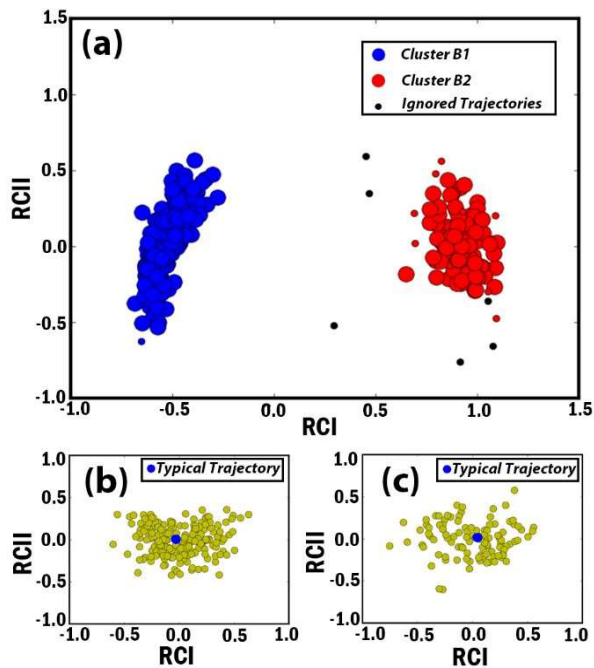

**Fig 5.** The further clustering analysis of all trajectories belonging to Cluster **B** when the propagation lasts to 1ps. (a) In the first run, we collected all trajectories in Cluster **B**, defined their pair-wise distance matrix, employed the dimensionality reduction approach and performed the trajectory clustering analysis. Two clusters appear, labelled as Cluster **B1** and Cluster **B2**. (b) In the second run, we took all trajectories belonging to Cluster **B1** and repeated the above analysis as the first step. (c) The similar analysis was performed for trajectories belonging to Cluster **B2**. The blue dots in (b) and (c) represent the typical trajectories of each cluster.

### 3.2.1 Geometrical Evolution of Trajectories belonging to Cluster B1

For all trajectories in Cluster **B1**, we checked their geometrical evolutions by the analysis of the geometry similarity. We took a snapshot at every 40 fs for each trajectory belonging to Cluster **B1** and 4775 geometries are collected.

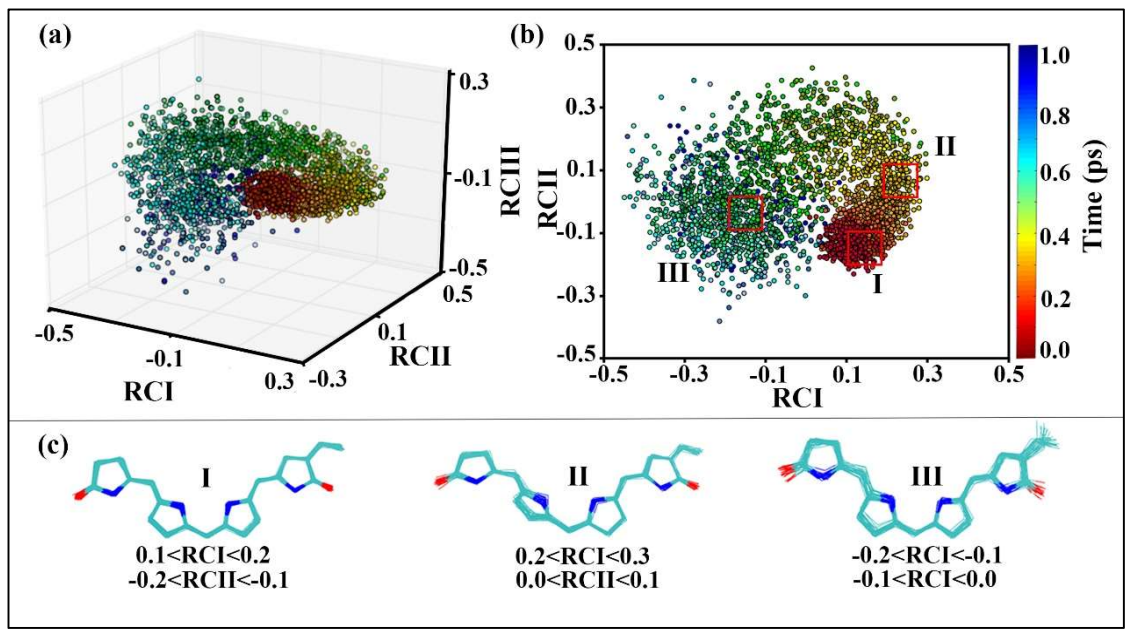

**Fig 6.** Classical MDS analysis of the geometrical evolution for the trajectories belonging to Cluster **B1**. (a) Locations of sampled geometries in the low-dimensional space spanned by three leading reduced coordinates and colour codes indicate the time. (b) Locations of sampled geometries in the low-dimensional space spanned by two leading reduced coordinates and three representative blocks. (c) Geometrical aggregations in three representative locations.

The classical MDS analysis of the geometry similarity over the trajectories belonging to Cluster **B1** gives a very interesting distribution pattern in the low-dimensional space shown in Fig 6 (a). Before ~ 500 fs, the first two leading dimensions control the propagations. After ~ 500 fs the third reduced coordinate starts to play an important role. We also show the data distribution and evolution in the two-dimensional reduced space in Fig 6 (b). Starting from the FC region (RCI ~ 0.15 and RCII ~ -0.15), the whole propagation seems to follow a circle. We selected three key blocks (I II III) in the representative regions (FC region, hopping region and photoproduct region) and aggregate all the snapshots in each block, shown in Fig 6 (c).

From Block I to Block II, the $\tau_{9\_10}$ twisting angle clearly experiences rotational motion accompanied by the weak torsional motion along $\tau_{5\_6}$. From Block II to Block

III both $\tau_{9\_10}$ and $\tau_{5\_6}$ angles return to the initial values while the "hot" geometries appear due to excessive energies. The large-amplitude vibrational motions not only include the $\tau_{9\_10}$ and $\tau_{5\_6}$ torsions, but also the vibrations of other coordinates such as the deformation of the side vinyl group. This explains that the third reduced coordinate RCIII is involved after the system goes back to the ground state. Also due to same reason, the data ensemble seems not finally go back to the starting region, while all photoproduct geometries look very similar to the initial reactant ones.

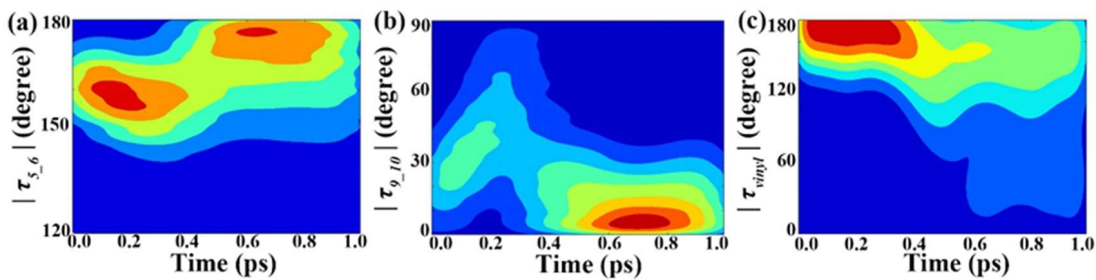

**Fig 7.** Time-dependent Distribution diagram of three key coordinates for the trajectories belonging to Cluster **B1**. (a) Distribution of $\tau_{5\_6}$ *vs* time. (b) Distribution of $\tau_{9\_10}$ *vs* time. (c) Distribution of torsion of vinyl group ($\tau_{vinyl}$) *vs* time.

To confirm the above analysis results, we gave the time-dependent distribution diagram of the $\tau_{9\_10}$, $\tau_{5\_6}$ and $\tau_{vinyl}$ group in Fig 7 (a), (b) and (c). Before 300 fs, the strong torsional motion of $\tau_{9\_10}$ is observed, accompanied with the weak change of $\tau_{5\_6}$. This observation is consistent with our previous theoretical results.[70] From 300 fs to 500 fs, both torsional angles return back to the initial configurations. After 500 fs, both $\tau_{9\_10}$ and $\tau_{5\_6}$ show the rather board distributions, also indicating their large-amplitude vibrational motions. The deformation of the side vinyl group starts to be very important after 400 fs, reflected by the evolution of $\tau_{vinyl}$. Most importantly, such motion is highly excited, because the distribution of $\tau_{vinyl}$ covers a very board angular range. Overall, we can still assign Cluster **B1** to be the channel, in which the system assesses the conical intersection by the strong $\tau_{9\_10}$ torsional motion and the weak $\tau_{5\_6}$ torsional motion, and then trajectories go back to the reactant region, if we only considered the backbone motion and neglect the side-chain motion.

## 3.2.2 Geometrical Evolution of Trajectories belonging to Cluster B2

Cluster **B2** contains 106 trajectories, which are examined by the same analysis protocol (based on 2650 geometries) as used in Cluster **B1**.

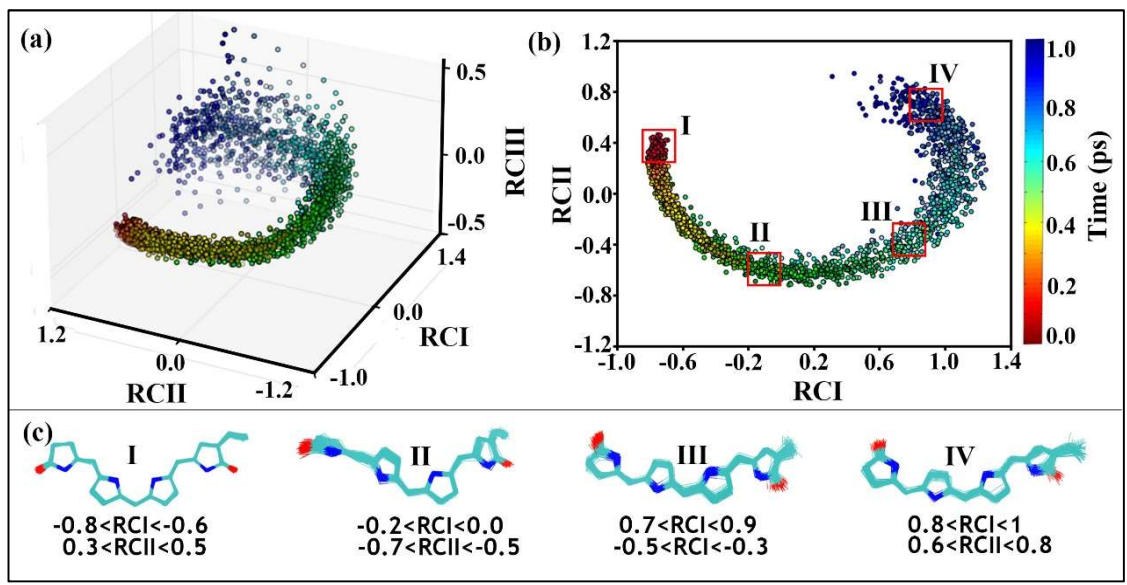

**Fig 8.** Classical MDS analysis of the geometrical evolution for the trajectories belonging to Cluster **B2**. (a) Locations of sampled geometries in the low-dimensional space spanned by three leading reduced coordinates and colour codes indicate the time. (b) Locations of sampled geometries in the low-dimensional space spanned by two leading reduced coordinates and four representative blocks. (c) Geometrical aggregations in four representative locations.

Compared to Cluster **B1**, the geometrical evolution of trajectories belonging to Cluster **B2** has a very clear propagation pattern as seen in Fig 8 (a) and (b), both in a three-dimensional space or a two-dimensional space. Before 600 fs the propagation is dominated by the first two key coordinates RCI and RCII. After that the third dimension RCIII starts to be involved. In the two-dimensional space spanned by RCI and RCII, as shown in Fig 8 (b), the geometry evolution basically follows a semi-circle, starting from the FC region. Then, we selected four representational blocks (I, II, III, IV) to examine the features of the geometry evolution In Fig 8 (c). From Block I (close to the FC region) to Block II, $\tau_{9\_10}$ increases from ~ 0° to ~ 90°, accompanied by the change of $\tau_{5\_6}$. From Block II to Block III, the $\tau_{9\_10}$ trends to be

~180° while the $\tau_{5\_6}$ returns to the initial values. In the later stage, $\tau_{14\_15}$ start to play roles from Block III to Block IV. At the same time, we also observe the large distribution of the $\tau_{vinyl}$ angle.

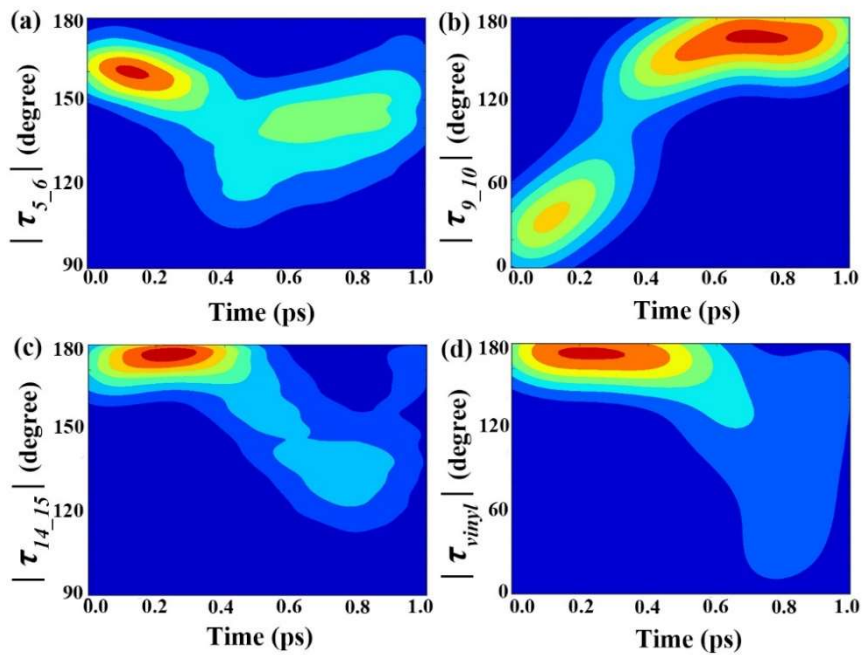

**Fig 9.** Time-dependent distribution diagram of four key coordinates for the trajectories belonging to Cluster **B**2. (a) Distribution of $\tau_{5\_6}$ *vs* time. (b) Distribution of $\tau_{9\_10}$ *vs* time. (c) Distribution of $\tau_{14\_15}$ *vs* time. (d) Distribution of $\tau_{vinyl}$ *vs* time.

We made the time-dependent distribution diagram of the four key torsion angles ($\tau_{5\_6}$, $\tau_{9\_10}$, $\tau_{14\_15}$ and $\tau_{vinyl}$) as shown in Fig 9. It is almost the same with our discussion on the geometry evolution. The $\tau_{9\_10}$ angle goes from ~ 0° to ~ 90°, and then continuously moves to ~ 180° to give the photoproducts before 500 fs. The $\tau_{5\_6}$ angle also displays the visible changes and then returns in the dynamics. Please notice that the $\tau_{14\_15}$ angle may also show some torsional motion. However, such motion starts to takes place only on the ground-state dynamics, even after the final products are almost formed and the $\tau_{14\_15}$ angle quickly goes back to the initial configuration as shown in Fig 9 (b) and (c). Thus, it is safe to believe that this angle is not relevant to the current analysis and no other isomer is formed by such motion. Similar to the cases in Cluster **B1**, we observe the large amplitude motion of the side vinyl group in Fig 9 (d). Overall, Cluster **B2** corresponds to the channel in which the system assesses the conical intersection by the $\tau_{9\_10}$ torsional motion and the weak $\tau_{5\_6}$ torsional motion, and then trajectories move towards to the photoproducts with $\tau_{9\_10} \sim 160° - 180°$.

**3.2.3 Typical Trajectory**

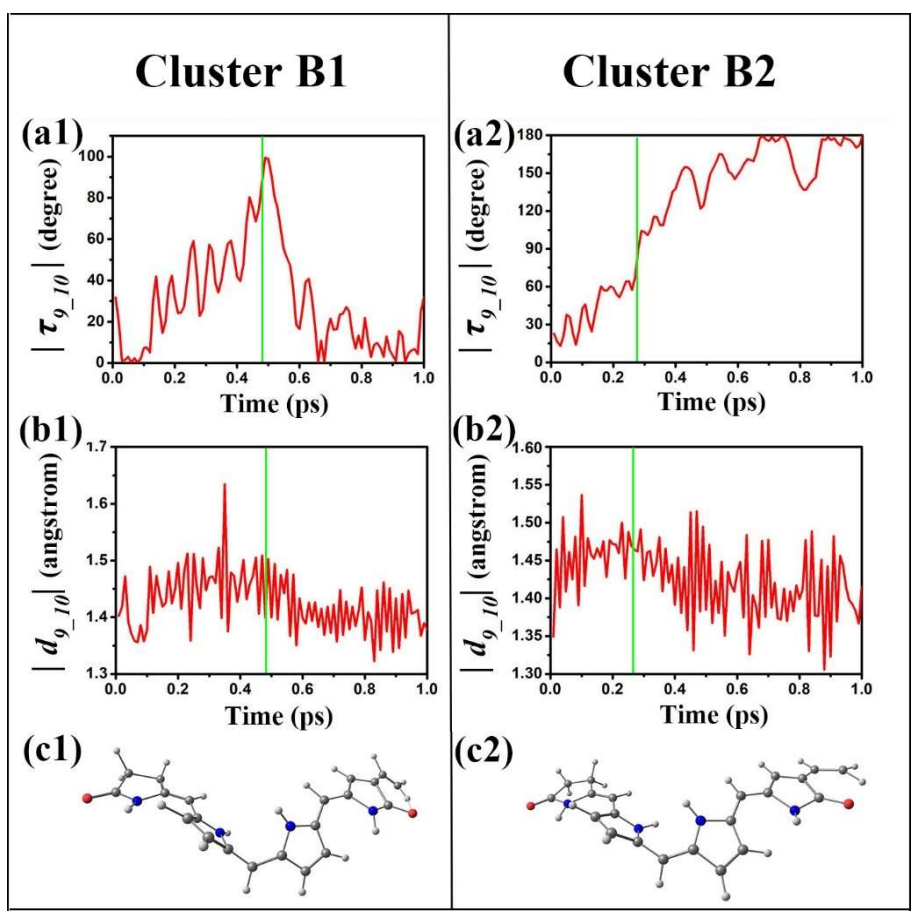

**Fig 10.** Geometrical evolution in typical trajectories for Cluster **B1** and **B2**. The time-dependent torsional angle $\tau_{9\_10}$, bond distance $d_{9\_10}$, the hopping geometries of the typical trajectory in the **B1** cluster are given in (a1), (b1), (c1) respectively. The corresponding results in the **B2** cluster are shown in (a2), (b2) and (c2) respectively. The vertical green lines represent the hopping events.

In the above analysis, we clearly demonstrated that it is possible to divide all trajectories into different clusters, while each cluster represents a reaction channel. In this situation, we can define the "typical trajectory" in each cluster. For Cluster **B1** and Cluster **B2**, their typical trajectories are given in Fig 10 (a1) - (c1) and (a2) - (c2), respectively. For illustration, we show a few important key coordinates *vs* time and give the evolution of other coordinates in Supporting Information. As shown in Figure 10 (a1) - (c1), in the typical trajectory that represents the evolution of Cluster **B1**, $\tau_{9\_10}$ increase from ~ 0° to ~ 100°, then return to ~ 0° in the dynamics. At 487 fs, the hop takes place with $\tau_{9\_10}$ ~ 90° in the vicinity of the conical intersection seam. It is also clear that the $C_9$-$C_{10}$ distance becomes longer in the early state of dynamics. All these features, including the time scale and the geometry evolution, are consistent with the above discussions. This strongly implies that a reasonable "representative" trajectory is selected. The same way can also be applied to select the typical trajectory for Cluster **B2**, see Figure 10 (a2) - (c2).

**3.3 Photoisomerization Mechanism, Reaction Channels and Branching Ratio**

Up to now, we employed the clustering analysis on the trajectory similarity to distinguish different channels in the nonadiabatic dynamics of the PΦB multi-channel photoisomerization. The configurational similarity analysis furtherly gives us the geometrical evolution feature of each channel. Most interestingly, the trajectory clustering analysis automatically provides a way to define the so-call typical trajectory.

Starting from the FC region, all trajectories are firstly grouped into two clusters, **A** and **B**, with the branching ratio around 0.31:0.66 (**A:B**). Then each cluster is again divided again according to their final products. At the end, four clusters are given, which are **A1**, **A2**, **B1** and **B2** with branching ratio 0.16:0.15:0.42:0.23. The sum of the total probability is not exact one, because some trajectories are neglected in clustering analysis. Clearly each cluster corresponds to a different reaction channel.

For Cluster **A1**, the system tends to follow the torsional motion along $\tau_{10\_11}$, performs the $S_1$-$S_0$ hops with $\tau_{10\_11}$ ~ 70°-90° near conical intersection, and then returns to reactants. Although some vibrational motions, such as the geometrical deformation of the side vinyl group, are excited, we still can attribute that this channel finally gives the reactants by checking the backbone motion. For Cluster **A2** the torsional motion of $\tau_{10\_11}$ is also responsible for the excited-state dynamics towards the conical intersections. After internal conversion, the trajectories tend to move forwards and to give the photoproducts with $\tau_{10\_11}$ ~ 160-180°.

For Cluster **B**, the system moves towards the conical intersection along the $\tau_{9\_10}$ torsional motion. After hopping back to the ground state, the trajectories belonging to Cluster **B1** return to the reactants, while trajectories belonging to Cluster **B2** continuously move to the photoproducts.

After all trajectories are clearly assigned into different clusters, we plot the important geometrical features in a few key events in the trajectory evolution associating to each cluster. For example, for each cluster, **A1**, **A2**, **B1** and **B2,** we show their hopping geometries and final products in Fig 11. Each cluster defines a reaction channel. This means that the current analysis can successively distinguish different reaction channels. Overall, all current results on the PΦB photoisomerization are highly consistent with our previous works.[70]

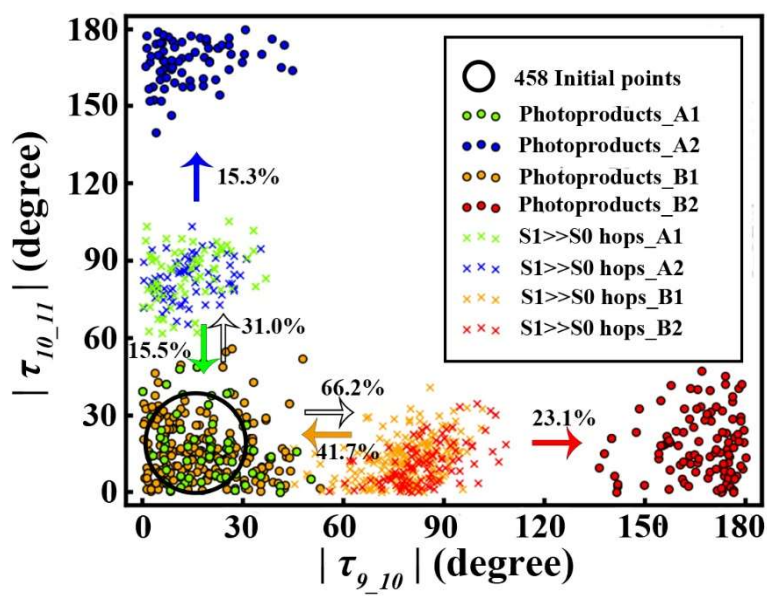

**Fig 11.** The branching ratios towards different channels in the TSH simulation of the PΦB's photoisomerization.

## 4. DISCUSSIONS

In the straightforward way to examine the on-the-fly TSH nonadiabatic dynamics, the evolution of each trajectory is examined one after another by eye view. It is also necessary to plot the hopping geometries, the final products, and the time-dependent evolution of relevant internal coordinates, to perform a meaningful analysis. Some preliminary knowledge on the possible reaction pathways and active coordinates is also necessary for the analysis task. Thus, when a large number of trajectories are employed, the analysis task becomes tedious even under the help of computational scripts. The current analysis approach, however, is more powerful, which automatically finds the reaction channels and branching ratio by the trajectory clustering analysis based on trajectory similarity. Then for each cluster, it is rather easy to extract the major geometry evolution responsible for the corresponding reaction channel.

It is also possible to perform the geometry similarity analysis in the configurational space directly, as shown in our previous works.[95] However, the current analysis based on both the trajectory similarity and the geometry similarity is somehow more powerful due to several reasons. First the dimensionality reduction approaches purely based on geometry similarity basically give a few of leading coordinates. This may not work well in the multi-channel situations. In the current approach, each cluster corresponds to a single reaction channel, thus all trajectories belonging to such cluster experiences the similar molecular motion. In this case, it is easier to get meaningful results because it is possible to perform the dimensionality reduction analysis for each single channel. This explains why sometimes a single reduced coordinate (even derived from the linear dimensionality reduction algorithm) may be good enough for the analysis of the geometry evolution. Second when a large number of trajectories are involved, the dimensionality reduction purely based on the geometry similarity requires the linear algebra operations on the extremely huge pair-wise dissimilarity matrix formed by a large number of geometries. This task may become very challenging because the calculation may require an extremely huge amount of computer memory to store and treat the very huge matrices. The current approach, on the other hand, requires smaller computer memory in the estimation of the trajectory similarity, although the total computational time should be slightly longer. When each cluster is identified, we only need to perform the dimensionality reduction analysis based on all trajectories belonging to the single cluster. Because the much smaller pair-wise dissimilarity matrix is considered in the dimensionality reduction approaches, the memory issue is largely alleviated. Third, we can also easily find the so-called "representative" trajectory for each channel from the current analysis. Overall, the current proposed analysis protocol is more powerful to analyze the multi-channel nonadiabatic dynamics.

In this work, we performed the trajectory clustering analysis in the two-step manner, namely first checking the responsible conical intersection and second examining the final products. In principle, it is always recommended to examining the excited-state dynamics before hops, because it is very important to understand the reaction channels of the excited-state dynamics and relevant conical intersections in the analysis of multi-channel nonadiabatic dynamics. In some cases, after the internal conversion the system may become highly vibrationally excited and the "hot" ground-state dynamics may not be very relevant to the nonadiabatic dynamics. In this case, only the first step in the current analysis protocol is necessary. Although it is possible to plot the hopping geometries in the examination of the reaction channels, the current analysis way displays many advantages, for instance taking the time evolution into account directly and giving us the representative trajectory for each channel. In addition, the excited-state motion is normally driven by a few of reactive coordinates in a single channel in the ultrafast nonadiabatic dynamics and the Fréchet distance may well capture the main geometrical evolution. As a contrast, sometimes the hot motion on the ground state may create many highly distorted snapshots even if the ground-state dynamics may follow some common pathways. In this case, the distance between two trajectories may be determined by these highly distorted geometries, instead of their different reaction channels via different conical intersections. Thus, it is more transparent to first check the relevant conical intersections and then the final products in more general cases.

## 5. CONCLUSION

We propose a powerful approach to analyze the TSH simulation results of the multi-channel nonadiabatic photoisomerization dynamics by considering both the trajectory similarity and the geometry similarity. In this approach, the clustering analysis of the trajectory similarity is first employed to find how many reaction channels are involved, while the active reaction coordinates responsible for each channel are then identified by the geometry similarity analysis in the configuration space without the requirement of the pre-known knowledge.

In practices, the analysis protocol starts from many trajectories obtained from TSH simulation. The trajectory similarity is estimated by their Fréchet distance. After the pair-wise Fréchet distance matrix was built for all trajectories, the DBSCAN clustering analysis is performed to assign trajectories into different groups. When each group cannot be divided any more, all trajectories belonging to the single non-separable cluster in principle are governed by the same individual reaction channel. To identify the major geometrical evolution feature in each reaction channel, we collect the geometries from the trajectories belonging to the same cluster and compute their pair-wise dissimilarity matrix. Then the MDS dimensionality reduction approach is performed to extract the major coordinates responsible for each channel. As a side product, it is very easy to find the so-called "representative" trajectory from this analysis protocol.

The multi-channel PΦB photoisomerization dynamics is used to explain this novel approach in this work. We first consider the excited-state dynamics and set the cutoff of trajectory propagation at hops. The clustering analysis of the trajectory similarity shows that two clusters are formed, which correspond two decay channels via their individual conical intersections. In the second step, we start from each cluster, take the photoreaction products into account and perform the same analysis again. At this step, we notice that the single cluster, obtained at the first step, can be divided into two clusters again. This means that after passing the same conical intersection the trajectories may go forwards to form the photoproduct or return to the reactant. Totally, four groups of trajectories can be clearly identified and each of them corresponds to a reaction channel. For all four reaction channels, it is possible to extract the active torsional motion and find the typical trajectory. All these results are consistent with our previous studies.[70]

This work demonstrates that the current analysis protocol can extract the main features of multi-channel nonadiabatic photoisomerization dynamics, such as the reaction channels, the branching ratio and relevant molecular motions, in a more automatic and intelligent way. This analysis approach should be very powerful, which can also be employed in other trajectory-based dynamics approaches.[11,17,23,33] The current work only focuses on the photoisomerization, while in principle the same approach can also be employed to treat more general types of nonadiabatic dynamics.[33,41,68] In more realistic cases, this analysis task may face additional problems, such as that the trajectory clustering analysis may not give the clearly-distinguishable cluster structure, or the estimation of geometry similarity may require more advanced molecular descriptors.[136-138] This represents an interesting challenging topic in future.

## ASSOCIATED CONTENT

**Supporting Information**

This Supporting Information is available free of charge on the ACS Publications website at http://pubs.acs.org.

Some figures relevant to Cluster **A**，time-dependent evolution of important coordinates in the typical trajectories in Cluster **B1** and Cluster **B2** (PDF)

## AUTHOR INFORMATION

**Corresponding Author**

*Fax: +86-532-80662778. Tel.: +86-532-80662630.

E-mail: lanzg@qibebt.ac.cn, zhenggang.lan@gmail.com

**Notes**

The authors declare no competing financial interest.

## ACKNOWLEDGMENT

This work is supported by NSFC Project (No. 21673266 and 21503248) and the Natural Science Foundation of Shandong Province for Distinguished Young Scholars (JQ201504). The authors thank Supercomputing Centre, Computer Network Information Center, CAS, National Supercomputing Center in Shenzhen, National Supercomputing Center in Guangzhou and the Super Computational Centre of CAS-QIBEBT for providing computational resources.

Supporting Information for

Analysis of Trajectory Similarity and Configuration Similarity in On-the-Fly Surface-Hopping Simulation on Multi-Channel Nonadiabatic Photoisomerization Dynamics

*Xusong Li,*[1,2,3] *Deping Hu,*[1,3] *Yu Xie,*[1] *and Zhenggang Lan*[1,2,a)]

[1]CAS Key Laboratory of Biobased Materials, Qingdao Institute of Bioenergy and Bioprocess Technology, Chinese Academy of Sciences, Qingdao 266101, China

[2]Sino-Danish Center for Education and Research/Sino-Danish College, University of Chinese Academy of Sciences, Beijing 100049, China

[3]University of Chinese Academy of Sciences, Beijing 100049, China

Email: lanzg@qibebt.ac.cn, zhenggang.lan@gmail.com

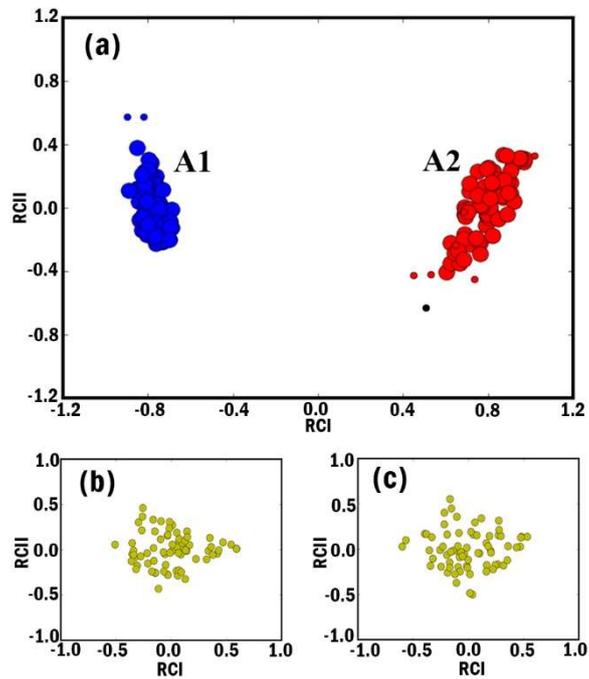

**Fig S1.** The further clustering analysis of all trajectories belong to the Cluster **A** when the propagation lasts to 1ps. (a) In the first run, we collected all trajectories in Cluster **A**, defined their pair-wise distance matrix, employed the dimensionality reduction approach and performed the trajectory clustering analysis. Two clusters appear, labelled as Cluster **A1** and Cluster **A2**. (b) In the second run, we took all trajectories belong to Cluster **A1**, defined the pair-wise distance matrix, employed the dimensionality reduction approach and performed the trajectory clustering analysis. (c) The similar analysis was performed for trajectories belong to Cluster **A2**.

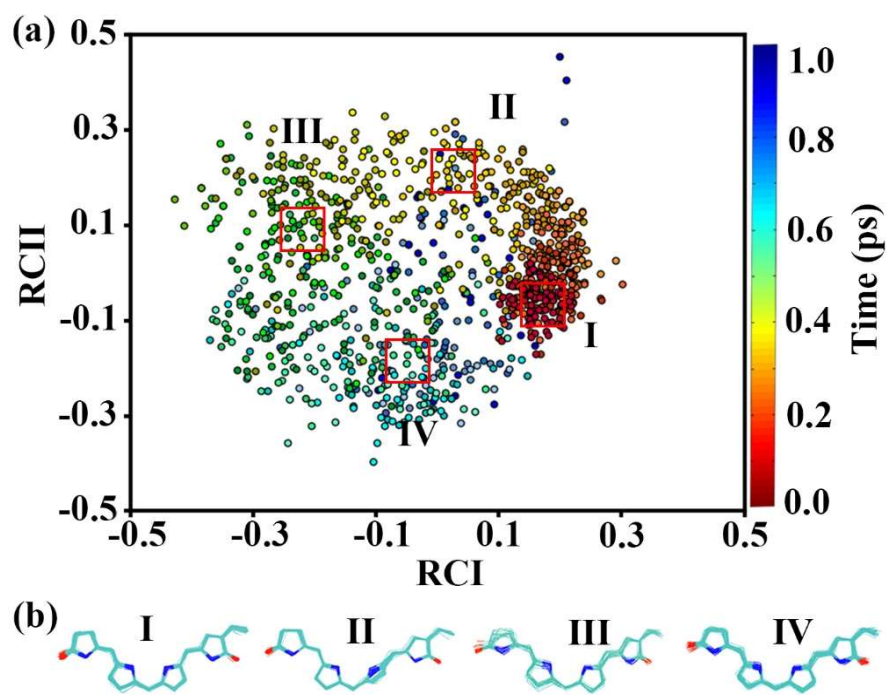

**Fig S2.** Classical MDS analysis of the nonadiabatic dynamics results of the trajectories belong to Cluster **A1**. (a) Locations of sampled geometries in the low-dimensional space spanned by two leading reduced coordinates and four representative blocks. (c) Geometrical aggregations in four representative locations.

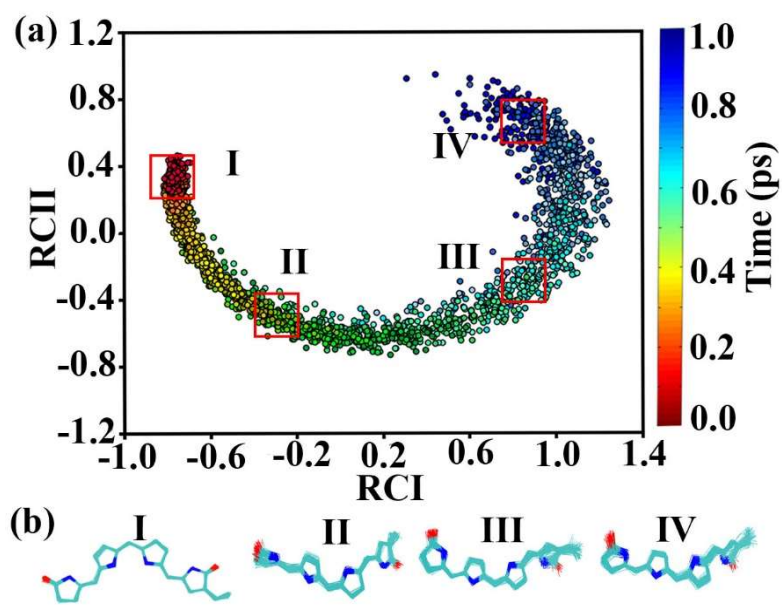

**Fig S3.** Classical MDS analysis of the nonadiabatic dynamics results of the trajectories belong to Cluster **A2**. (a) Locations of sampled geometries in the low-dimensional space spanned by two leading reduced coordinates and four representative blocks. (c) Geometrical aggregations in four representative locations.

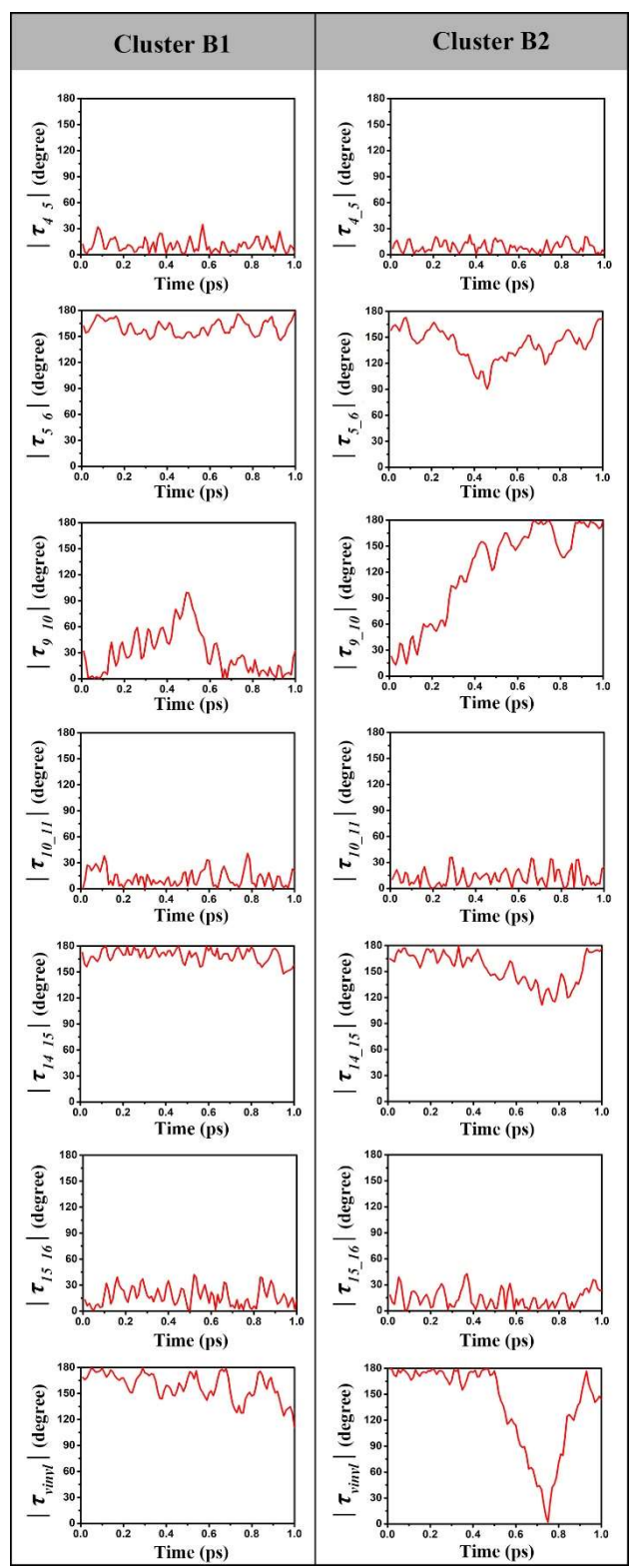

**Fig S4.** The propagation of seven key torsion angles *vs* time in the typical trajectories in Cluster **B1** and **B2**. Please notice that $\tau_{14\_15}$ may also show some torsional motion. However, such motion starts to takes place only on the ground-state dynamics, even after the final products are almost formed. In addition, the angle $\tau_{14\_15}$ quickly goes back to the initial configuration. Thus it is safe to believe that this angle is not relevant to the current analysis and no other isomer is formed by such motion.

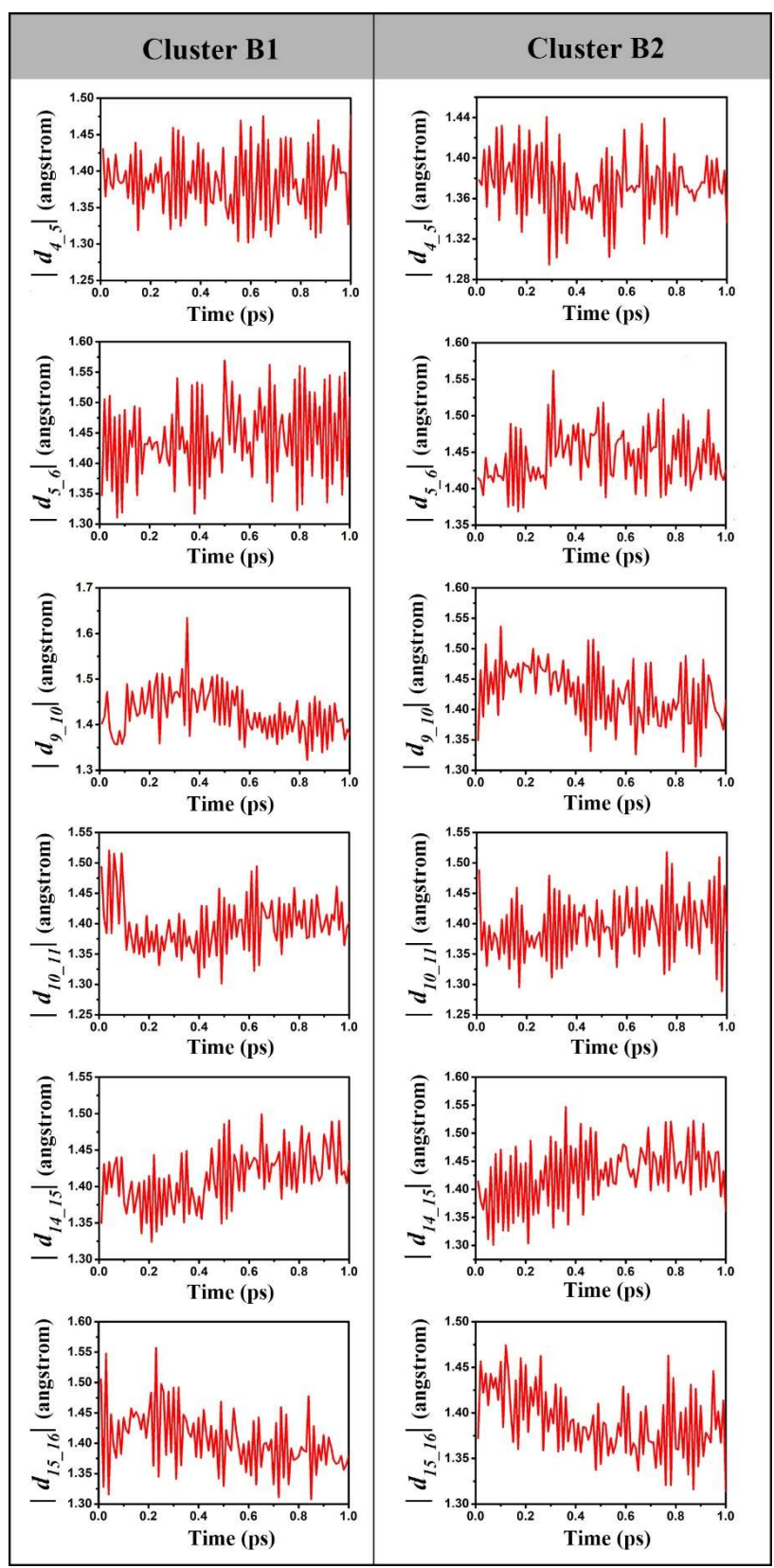

**Fig S5.** The propagation of six key bond distance *vs* time in the typical trajectories in Cluster **B1** and **B2**.